\documentclass[prb,twocolumn,showpacs,preprintnumbers,amsmath,amssymb]{revtex4}

\usepackage{graphicx}
\usepackage{dcolumn}
\usepackage{bm}

\begin{document}

\title{Voltage-induced Shapiro steps in a superconducting multi-terminal structure}

\author{J.C. Cuevas$^{1,2,3}$ and H. Pothier$^{4}$}

\affiliation{$^1$Departamento de F\'{\i}sica Te\'orica de la Materia Condensada, 
Universidad Aut\'onoma de Madrid, E-28049 Madrid, Spain} 
\affiliation{$^2$Institut f\"ur Theoretische Festk\"orperphysik, Universit\"at Karlsruhe,
D-76128 Karlsruhe, Germany} 
\affiliation{$^3$Forschungszentrum Karlsruhe, Institut f\"ur Nanotechnologie, D-76021
Karlsruhe, Germany} 
\affiliation{$^4$Quantronics Group, Service de Physique de l'\'Etat Condens\'e, 
DRECAM, CEA-Saclay, 91191 Gif-sur-Yvette, France}

\date{\today}

\begin{abstract}
When a superconducting tunnel junction at a finite voltage is irradiated with
microwaves, the interplay between the alternating Josephson current and the ac
field gives rise to steps in the dc current known as Shapiro steps. In this
work we predict that in a mesoscopic structure connected to several
superconducting terminals one can induce Shapiro-like steps in the absence
of any external radiation simply by tuning the voltages of the leads. To
illustrate this effect we make quantitative predictions for a three-terminal 
structure which comprises a diffusive superconductor-normal metal-superconductor 
(SNS) junction and a tunneling probe, a set-up which can be realized experimentally.
\end{abstract}


\maketitle

\section{Introduction}

In 1962 Josephson predicted that a tunnel junction between two superconductors
at a finite bias $V$ would sustain an alternating current with frequency
$2eV/\hbar$.~\cite{Josephson1962} Josephson also argued that the alternating
current would be frequency-modulated by an applied rf field and this would
lead to steps in the dc current-voltage characteristics at bias
voltages given by $V = m h \nu/2e$, where $\nu$ is the frequency
of the field and $m$ is an integer. In 1963 Shapiro shone microwaves onto
a junction and observed the predicted steps,~\cite{Shapiro1963} which since
then are referred to as \emph{Shapiro steps}. This experiment constituted
the first proof of ac Josephson effect and also paved the way for 
applications of this effect.~\cite{Barone1982}

The Shapiro steps are a consequence of the interplay between an ac Josephson 
current and a microwave signal. The observation of Shapiro steps does not 
necessarily require an external microwave generator: when two Josephson 
junctions are electromagnetically coupled like in the classical experiment of 
Giaever,~\cite{Giaever1965} a voltage-biased junction can be used as a 
microwave source and induce Shapiro steps in the current-voltage characteristics
(I-V) of the other one. More subtle coupling schemes have been achieved, 
in which two superconducting weak links placed a few micrometers apart 
share a common electrode. Shapiro-like steps were then observed in the I-Vs, 
which were attributed to the phase-locking of the Josephson frequencies 
when the dc voltages on two bridges are approximately matched.~\cite{Lindelof1977} 
Although there is no rigorous microscopic theory of this non-equilibrium effect, 
it is usually attributed to the diffusion of the quasiparticle charge imbalance 
generated in and around the weak links.~\cite{Hansen1984} Very recently, it
has been predicted that one could also generate Shapiro steps by coupling a tunnel
Josephson junction to a mechanical oscillator.~\cite{Zhu2006}

\begin{figure}[t]
\begin{center}
\includegraphics[width=0.7\columnwidth,clip]{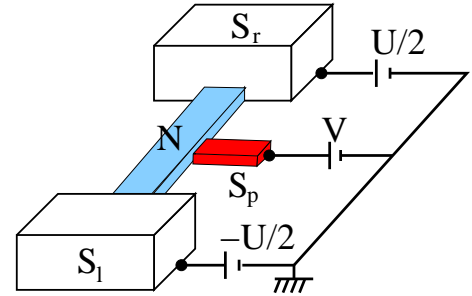}
\caption{\label{setup} (Color online) Schematic representation of the setup:
a metallic wire (N) is connected at its ends to superconducting reservoirs
S$_l$ and S$_r$, biased at potentials $-U/2$ and $U/2$, respectively.
An additional superconducting probe (S$_p$) at a potential
V is attached to the wire through a tunnel junction.}
\end{center}
\end{figure}

In this work we predict that a different type of Shapiro-like steps can appear in a 
single multi-terminal superconducting structure without any external radiation. The 
idea goes as follows. Inspired by the experiment of Ref.~[\onlinecite{Pierre2001}], 
let us consider the three-terminal structure shown in Fig.~\ref{setup}. It is formed 
by a diffusive SNS junction, and a superconducting tunneling probe S$_p$ 
attached to the normal wire. If one applies a potential difference $U$ across
the wire, ac currents with a frequency $2eU/h$ and its harmonics will flow 
along the N wire. If a potential $V$ is now applied to the probe electrode, it 
will generate ac currents with Josephson frequencies $2e(V\pm U/2)/h$. The
interference between these ac currents which both involve the quasiparticles in 
the diffusive wire gives rise to contributions to the dc current through the 
interface N-S$_p$, which in this setup appear as Shapiro steps at discrete 
voltages $V = m U/2$, where $m$ is an odd integer. Notice that in this case
the steps appear due to the non-equilibrium state created in the diffusive
wire. In this region the energy distribution function of quasiparticles is 
driven to oscillate at different frequencies by the dc voltages of the leads 
and the Shapiro steps are a result of the beating of these internal Josephson 
frequencies. As illustrated below, this effect is described quantitatively 
using the quasiclassical theory of superconductivity for diffusive systems, in 
particular time-dependent Usadel equations. Finally, it is important to emphasize 
that these \emph{voltage-induced Shapiro steps} may appear in a great variety 
of mesoscopic structures and, as explained below, it constitutes a valuable way 
to directly test the ac Josephson effect.

The rest of the paper is organized as follows. In section II we remind the basic
physics of diffusive SNS junctions at finite bias. Section III is devoted to
the description of the theoretical formalism and contains the main technical
results. In section IV we discuss the numerical results for both the current-voltage
characteristics of the tunneling probe and the dependence of the height of the 
voltage-induced Shapiro steps on the different system parameters. Finally, in
section V we briefly summarize the main conclusions of this work.

\section{Basic physics of diffusive SNS junctions}

The effect described above is a consequence of the non-equilibrium
properties of diffusive SNS junctions. The physics of these
junctions is the result of the interplay between the \emph{proximity effect} in the
normal wire and the occurrence of \emph{multiple Andreev reflections} (MARs).
The proximity effect is the modification of the properties of a normal metal in
contact with a superconductor and it has been extensively studied in
diffusive hybrid nanostructures.~\cite{Pannetier2000} MARs are the tunneling
processes that dominate the subgap transport in SNS junctions. Here, successive 
Andreev reflections at both S electrodes lead to a progressive rise of the 
quasiparticle energy until its energy exceeds the gap energy, 
$\Delta$.~\cite{Klapwijk1982} The interplay between proximity effect and MARs in 
diffusive SNS junctions gives rise to a rich variety of phenomena. Thus for 
instance, the conductance exhibits a very peculiar subgap structure.~\cite{Kutchinsky1997}
Dubos \emph{et al.}~\cite{Dubos2001b} studied the ac Josephson effect by shining
microwaves onto a junction, and they observed the appearance of fractional Shapiro
steps, which is a signature of a non-sinusoidal supercurrent-phase relation.
On the other hand, using the setup of Fig.~\ref{setup} Pierre
\emph{et al.}~\cite{Pierre2001} measured the non-equilibrium distribution function
in a long silver wire, where the proximity effect was negligible. They showed that
this function exhibits several steps, which is a manifestation of MARs.

The consequences of the interplay between proximity effect and coherent MARs in the
dissipative current has been addressed theoretically only very recently by using
the time-dependent Usadel equations.~\cite{Cuevas2006} This theory describes the
crossover from the short junction regime ($L \ll \xi=\sqrt{\hbar D/
\Delta}$),~\cite{Bardas1997} where $L$ is the wire length, $\xi$ is the superconducting
coherence length and $D$ the diffusion constant, to the incoherent limit ($L \gg
\xi$).~\cite{Bezuglyi2000} The intermediate regime is the relevant one for
the observation of the voltage-induced Shapiro steps.

\section{Calculation of the voltage-induced Shapiro steps}

We consider the structure depicted in Fig.~\ref{setup}, where the position of the 
tunneling probe along the wire is denoted by $x$ ($x_l=0$ and $x_r=L$). We assume 
that the three superconductors (S$_l$, S$_r$ and S$_p$) have the same energy gap $\Delta$.
Our goal is the calculation of the current $I(V)$ through the probe junction S$_p$
when a voltage $U$ is applied across the wire. We assume the N-S$_{l,r}$ interfaces
to be fully transparent and neglect phase-breaking phenomena.

To solve our problem we use the quasiclassical theory of superconductivity for
diffusive systems.~\cite{Usadel1970} This theory is formulated in terms of
momentum averaged Green functions ${\bf \check G}({\bf R}, t, t^{\prime})$
which depend on position ${\bf R}$ and two time arguments. These propagators
are $2\times 2$ matrices in Keldysh space ($\check \; $), where each entry is
a $2\times 2$ matrix in electron-hole space ($\hat \; $):
\begin{equation}
\label{keldysh-space}
{\bf \check G} = \left( \begin{array}{cc}
\hat G^R & \hat G^K \\
   0     & \hat G^A
\end{array} \right); \hspace{5mm}
\hat G^{R} = \left( \begin{array}{cc}
{\cal G}^{R} & {\cal F}^{R} \\
\tilde {\cal F}^{R}  & \tilde {\cal G}^{R}
   \end{array} \right) .
\end{equation}
\noindent
The Green functions for the left (l) and right (r) leads and for the probe (p)
electrode can be written as ${\bf \check G}_{j}(t,t^{\prime}) =
e^{-i \phi_j(t) \hat \tau_3/2\hbar} {\bf \check G}_0(t-t^{\prime})
e^{i \phi_j(t^\prime) \hat \tau_3/2 \hbar}$, where $\phi_j(t)$ is the phase
of the order parameter of the electrode $j=l,r,p$ given by
$\partial \phi_j(t) / \partial t = 2\mu_j/\hbar$, where the chemical
potentials are $\mu_l=-eU/2$, $\mu_r=eU/2$ and $\mu_p=eV$. Here,
${\bf \check G}_0(t)$ is the equilibrium bulk Green function of a BCS
superconductor. We now transform to energy representation, in which the
propagator ${\bf \check G}({\bf R},\epsilon, \epsilon^{\prime})$ depends
on two energy arguments. It satisfies the non-stationary Usadel equation,
which in the N region reads
\begin{equation}
\label{usadel-eq}
 \frac{\hbar D}{\pi} \nabla \left( {\bf \check G} \circ \nabla {\bf \check G}
 \right) + \epsilon \hat{\tau}_3  {\bf \check G} - {\bf \check G} \hat{\tau}_3
 \epsilon^{\prime} = 0 ,
\end{equation}
\noindent
where $\hat{\tau}_3$ is the Pauli matrix in electron-hole space. The convolution
product $\circ$ is defined as $({\bf \check A} \circ {\bf \check B})
(\epsilon,\epsilon^{\prime}) = \int d\epsilon_1 \; {\bf \check A}(\epsilon,
\epsilon_1) {\bf \check B}(\epsilon_1, \epsilon^{\prime})$. Equation
(\ref{usadel-eq}) is supplemented by the normalization condition ${\bf \check G}
\circ {\bf \check G} = -\pi^2 \delta(\epsilon - \epsilon^{\prime}){\bf \check 1}$.
Due to the finite bias $U$, to solve Eq.~(\ref{usadel-eq}) is a formidable task,
which is explained in detail in Ref.~[\onlinecite{Cuevas2006}]. What
matters for our discussion is that the Green functions in the wire adopt the form:
\begin{equation}
\label{g-time}
{\bf \check G}({\bf R},\epsilon,\epsilon^{\prime}) = \sum_m {\bf \check G}_{0,m}(\epsilon)
\delta(\epsilon_m - \epsilon^{\prime}) ,
\end{equation}
\noindent
where $\epsilon_m = \epsilon + meU$. Here, $m$ is an even integer for the diagonal 
components of the Green functions in Nambu space and an odd integer for the 
off-diagonal ones. Eq.~(\ref{g-time}) means, in particular, that the energy 
distribution function and the density of states in the normal wire oscillates 
with the Josepshon frequency $2eU/h$ and its harmonics. These oscillations result 
in a parametric pumping of the N-S$_{p}$ junction and hence to Shapiro steps.

Assuming that the electrode S$_p$ is weakly coupled to the N wire, we can express
the time-dependent current $I(V,t)$ through the tunnel probe up to first order
in the tunneling conductance $G_T$ as follows~\cite{Kuprianov1988}
\begin{equation}
\label{tunnel-current}
I(V,t) = \left( \frac{G_T}{8\pi e} \right) \int dt_1 \mbox{Tr} \left\{ \hat \tau_3
\left[ {\bf \check G}_w (t,t_1), {\bf \check G}_p(t_1,t) \right]^K \right\} ,
\end{equation}
\noindent
where ${\bf \check G}_w$ is the Green function of the wire at the position of the probe
junction and ${\bf \check G}_p$ is the Green function of the probe electrode, i.e.
a bulk BCS Green funtion. Using Eqs.~(\ref{g-time},\ref{tunnel-current}) it is easy
to show that
\begin{equation}
\label{harmonics}
I(V,t) = \sum^{\infty}_{m=-\infty} \sum^{1}_{n=-1} I^m_n(V)
e^{i \left( n\phi + 2neVt/\hbar + meUt/\hbar \right) } ,
\end{equation}
\noindent
where $\phi$ the dc part of the phase difference. From Eq.~(\ref{harmonics}),
one can distinguish two contributions to the dc current, which we shall simply
denote as $I$. First, there is a background current $I_B=I^0_0$ that contributes
at every voltage $V$. More important, there is a series of contributions at
discrete voltages $V_m = mU/2$ with $m$ odd,~\cite{note0} which are given by
$I_{\rm Shapiro}(\phi) = \sum_{m>0} (I^{-m}_1 e^{i\phi} + I^m_{-1} e^{-i\phi})
\delta(V-V_m)$, where $I^{-m}_1 = (I^m_{-1})^*$. Thus, \emph{Shapiro-like
resonances are induced every time the probe voltage $V$ is an odd multiple of
$U/2$}. From Eq.~(\ref{tunnel-current}), one can show that the ac current components
in Eq.~(\ref{harmonics}) can be expressed as
\begin{eqnarray}
\label{ac-components}
I^{m}_{1} & = & \left( \frac{G_T}{8\pi^2 e} \right) \int^{\infty}_{-\infty} d\epsilon \;
\left\{ \left[ {\cal F}^{R}_w \right]_{0,m}(\epsilon)
\tilde {\cal F}^{K}_{p}(\epsilon_m+eV) \right.  \nonumber \\
& & \hspace*{2.0cm} \left.  + \left[ {\cal F}^{K}_w \right]_{0,m}(\epsilon)
\tilde {\cal F}^{A}_{p}(\epsilon_m+eV) \right\} .
\end{eqnarray}
\noindent
Here, $\left[ {\cal F}^{R,K}_w \right]_{0,m}(\epsilon)$ are Fourier components of
the anomalous Green functions induced in the normal wire by proximity effect.
These components contain the information of ac Josephson effect in the SNS junction
and depend on $L$, $U$ and $x$. Obviously, when $L \gg \xi$ these components vanish
and in turn the Shapiro steps.~\cite{note1} The functions $\tilde {\cal F}^{A,K}_{p}
(\epsilon)$ are the bulk Green functions of the electrode S$_p$ given by $\tilde
{\cal F}^{A}_{p}(\epsilon) = -\pi \Delta / \sqrt{\Delta^2 - (\epsilon -i0^+)^2}$ and
$\tilde {\cal F}^{K}_{p}(\epsilon) = -2i \mbox{Im} \{ \tilde {\cal F}^{A}_{p}(\epsilon)
\} \tanh(\beta \epsilon /2)$, where $\beta=1/k_B T$ is the inverse of the temperature.

For completeness let us also mention that the background current can be
approximated by
\begin{eqnarray}
\label{back-aprox}
I_B & = & \left( \frac{G_T}{e} \right) \int^{\infty}_{-\infty} d\epsilon \;
\rho_p(\epsilon-eV) \rho_w(\epsilon) \left[ f(\epsilon-eV) - f_w(\epsilon) \right],
\nonumber
\end{eqnarray}
where $f(\epsilon)$ is the Fermi function, $f_w(\epsilon)$ is the dc part of
the distribution function of the N wire at the position of $S_p$,
$\rho_p(\epsilon)$ is the bulk BCS density of states and $\rho_w(\epsilon,U) =
\mbox{Im} \{ \left[ {\cal G}^{A}_w \right]_{0,0} \} /\pi$ is the non-equilibrium
spectral density in the normal wire.

\section{Results and discussion}

\begin{figure}[t]
\begin{center}
\includegraphics[width=\columnwidth,clip]{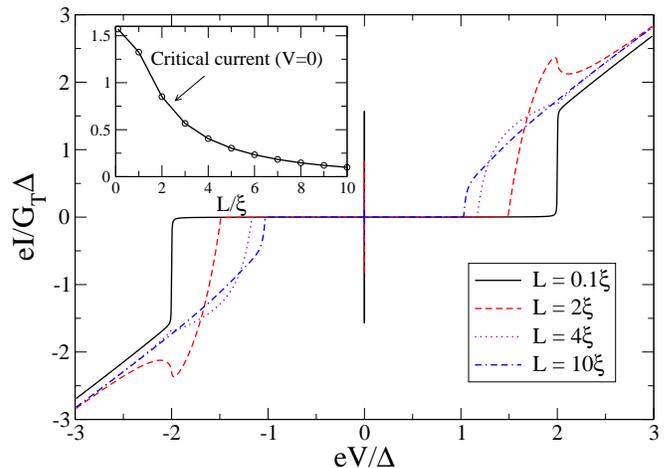}
\caption{\label{zero-bias} (Color online) Zero-temperature dc current in the probe
electrode S$_p$ at zero bias ($U=0$) for different lengths. The probe is located
on $x=L/2$. The inset shows the critical current as a function of the wire length.}
\end{center}
\end{figure}

It is illustrative to start the discussion of the results with the
case of zero-bias ($U=0$). In this limit there are no Shapiro steps, but the
dc current exhibits a supercurrent peak at V=0. This can be seen in
Fig.~\ref{zero-bias}, where we plot the dc current through a probe located on
$x=L/2$ for different wire lengths.~\cite{note2} Apart from the supercurrent, the
main feature of these I-V curves is the presence of a gap, which is
equal to $\Delta+ \Delta_g$, where $\Delta_g$ is the well-known minigap present
in the N wire due to proximity effect (see for instance Fig. 1 in
Ref.~\onlinecite{Cuevas2006}). Such a minigap decays with $L$ as $\Delta_g \sim
3.2\epsilon_T$, where $\epsilon_T = \hbar D/L^2$ is the Thouless energy. In the
inset of Fig.~\ref{zero-bias} we show how the critical current decays with $L$.
This critical current is a good reference for the height of the Shapiro steps at
finite $U$. Notice in particular, that for $L \ll \xi$ the critical current adopts
the value $I = \pi G_T \Delta /2e$, which is simply the critical current of a tunnel
junction between BCS superconductors.~\cite{Ambegaokar1963}

\begin{figure}[t]
\begin{center}
\includegraphics[width=\columnwidth,clip]{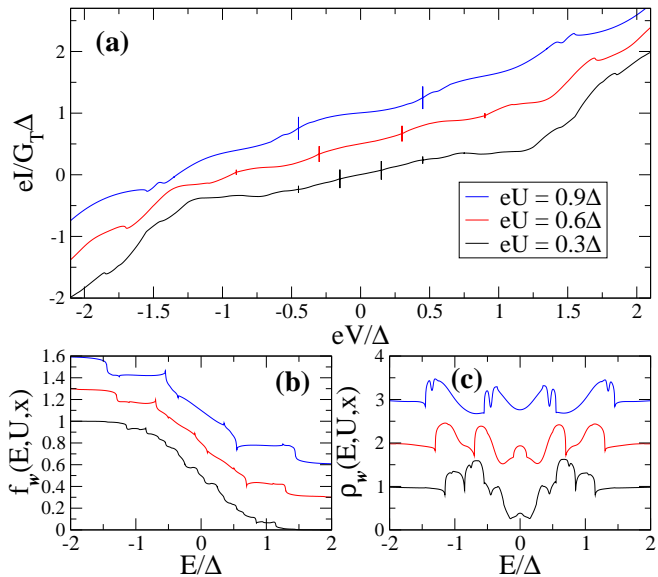}
\caption{\label{dc-current} (Color online) 
(a) Zero-temperature dc current in the probe S$_p$ located on $x=L/2$
for $L=2\xi$ and three different values of the voltage $U$: from bottom to top
$eU/\Delta=0.3,0.6,0.9$. (b) dc part of the distribution function of the normal wire
at $x=L/2$ as a function of the energy for the three voltages above. (c) Corresponding
non-equilibrium spectral density in the normal wire ($x=L/2$) as a function of the energy.
In the three panels the two upper curves have been shifted upwards for clarity.}
\end{center}
\end{figure}

Now, we turn to the case of finite bias $U$. To give a first 
impression of how the steps would appear in an experiment, we show in 
Fig.~\ref{dc-current}(a) the total contribution to the dc current, i.e.
background current plus Shapiro steps, for $L=2\xi$, $x=L/2$ and three different 
values of the voltage $U$. Notice that since we are assuming that the system is 
voltage biased, the Shapiro steps appear as peaks in the current, rather 
than steps as in a current-biased contact.~\cite{Shapiro1963} We have also plotted 
in Fig.~\ref{dc-current} both the distribution function and the non-equilibrium spectral 
function, which determine the shape of the background current. The first two peaks in the I-V 
at $V=\pm U/2$ could simply be viewed as the supercurrent peaks that correspond to 
the condition of equality between the potentials of the probe S$_{p}$ and of
one of the reservoirs S$_{l,r}$. However, their height depends on $U$ and they 
are smaller than the critical current at $U=0$ because the quasiparticles in the wire 
are out of equilibrium. Notice, also, the large sub-gap current, which is absent 
in Fig.~\ref{zero-bias}. More obvious manifestations of the interference of ac 
Josephson currents are the higher-order Shapiro steps, which appear at voltages 
that do not correspond to the alignment of the potential of S$_{p}$ to that of a 
reservoir. In Fig.~\ref{dc-current}(a), they are clearly seen at $V=\pm 3U/2$ at 
the lowest voltage ($U=0.3 \Delta$), but they progressively disappear as $U$ increases.

\begin{figure}[t]
\begin{center}
\includegraphics[width=\columnwidth,clip]{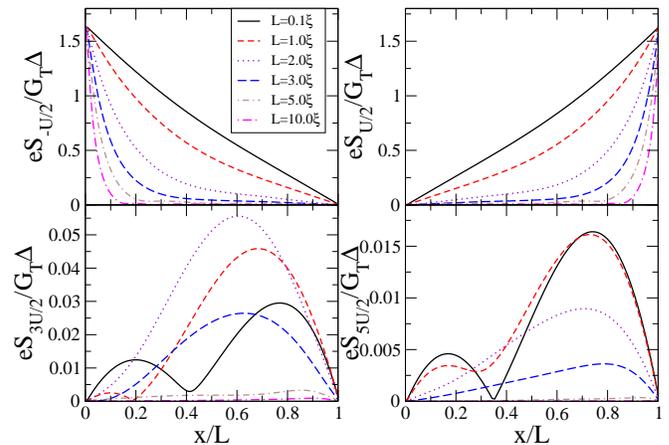}
\caption{\label{space} (Color online) Height of the Shapiro steps at zero
temperature as a function of the position of the probe ($x$) for $eU=0.5\Delta$
and different wire lengths.}
\end{center}
\end{figure}

Let us now study systematically the best conditions for the observation of the
voltage-induced steps. First, we analyze their dependence on the wire length
and the position of the probe at a given voltage $U$.
In Fig.~\ref{space} we show for $eU=0.5\Delta$ the height of the steps, denoted
as $S_{mU/2}$ for a step at $V=mU/2$, as a function of $x$ for different wire
lengths. Notice that we show the steps for positive voltages $V$ and the first
one for negative voltages, $S_{-U/2}$. Indeed, one can show that in this setup
the following relation holds: $S_{-mU/2}(x/L) = S_{mU/2}(1-x/L)$ for $m > 0$,
which can be seen in the upper panels of Fig.~\ref{space}. The most
remarkable features of Fig.~\ref{space} are the following. First, the step
$S_{U/2}$ is of the same order as the critical current at $U=0$, although a
bit smaller. Second, due to the biasing in this setup (see Fig.~\ref{setup}), the
step $S_{U/2}$ vanishes close the left electrode. The same happens with the steps
$S_{3U/2}, S_{5U/2},...$ at both boundaries because we are assuming that the
electrodes are ideal bulk reservoirs. Third, the higher-order steps ($m > 1$)
are, for this particular voltage $U$, at least an order of magnitude smaller than
$S_{U/2}$. Finally, with respect to the length dependence, the steps progressively
disappear as $L$ increases. For a given position, they decay roughly as
$\xi/L$ for $L \gg \xi$, which is the usual decay of induced superconducting
correlations in a normal diffusive system. However, for $L$ of the order of
$\xi$, the dependence is not neccesarily monotonous, as it can be
seen for the steps $S_{3U/2}$ and $S_{5U/2}$.

\begin{figure}[t]
\begin{center}
\includegraphics[width=\columnwidth,clip]{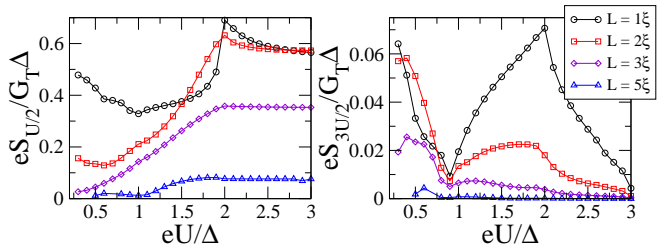}
\caption{\label{voltage} (Color online) Height of the first two Shapiro steps
for positive voltages in $x=L/2$ at zero temperature as a function of $U$ for
different lengths $L$.}
\end{center}
\end{figure}

The other crucial parameter that controls the height of the steps is the voltage 
$U$. As Eq.~(\ref{ac-components}) shows, in order to have a step at $V=mU/2$ ($m$ odd), 
one needs to have a non-zero component $\left[ {\cal F}^{R,K}_w \right]_{0,m}$. Such an ac 
component is related to the coherent transfer of $n=(|m|+1)/2$ Cooper pairs through 
the SNS junction, which requires the occurrence of coherent MARs of at least order $n$. 
An $n$-order process contributes significantly for voltages $2\Delta/n < eU < 2\Delta/(n-1)$ 
and its probability decreases as the voltage increases. Thus, one naively expects (i) 
high-order ($m>1$) steps to be more clearly visible at low bias $U$ and that (ii) at
voltages $U > 2\Delta$, only the steps $S_{\pm U/2}$ survive. Indeed, these expectations 
are confirmed by the calculations, as can be seen in Fig.~\ref{voltage}, where we show 
the height of the first two steps (for positive voltages) as a function of $U$ for 
different wire lengths and a probe located on $x=L/2$. Notice that the step
$S_{U/2}$ reaches its maximum at $\sim 2\Delta$ and then decays very slowly as the voltage increases whereas $S_{3U/2}$ vanishes 
rapidly for voltages above the gap.

\section{Conclusions}

In summary, we have predicted the possibility to generate parametrically
Shapiro steps in multi-terminal superconducting structures by tuning the
voltages in the reservoirs. We have illustrated this effect with detailed
calculations for the case a diffusive SNS system with a third electrode 
coupled to the N region through a tunnel junction. Our results, based on
the solution of the time-dependent Usadel equations, show that these 
voltage-induced steps are a direct manifestation of the ac Josephson effect 
in the SNS junction and they are visible as long as there are sizable 
superconducting correlations induced in the N wire. We stress that
these steps can be induced in any coherent mesoscopic structure attached to 
several (more than two) superconducting electrodes. Moreover, if in the setup of
Fig.~\ref{setup} all the electrodes were strongly coupled to the central wire, 
new phenomena would occur like the appearance of fractional Shapiro steps for 
$V=(m/n)U/2$, with $m,n$ integers. Thus, the effect predicted here suggests the
appearance of new physical phenonema and in particular, provides a new
way to directly test the ac Josepshon effect in mesoscopic structures.

\acknowledgments

We acknowledge fruitful discussions with S. Bergeret, W. Belzig, D. Esteve, J.C. Hammer, 
A. Levy Yeyati, A. Mart\'{\i}n-Rodero and C. Urbina. This work has been financed 
by the Spanish CYCIT (contract FIS2005-06255) and by the Helmholtz Gemeinschaft
(contract VH-NG-029).



\begin{thebibliography}{}
\bibitem{note0}
 One can show from Eq.~(\ref{tunnel-current}) that only those ac components that
 involve the anomalous Green functions of the wire can give rise to Shapiro
 steps, i.e. components with an odd index $m$.
\bibitem{note2}
 In Fig.~\ref{zero-bias} the phase difference between S$_l$ and S$_r$ is set
 to zero, which corresponds to the case where the supercurrent through
 the probe S$_p$ reaches its maximum.
\end{thebibliography}
\end{document}